\documentclass[reprint, amsmath, amssymb, aps]{revtex4-2}

\usepackage{times}
\usepackage{graphicx}   % Include figure files
\usepackage{bm}         % bold math
\usepackage{xcolor}
\usepackage[colorlinks=true,allcolors=blue]{hyperref} % hyperlinks
\usepackage{dsfont}     % for \mathds
\let\vec\boldsymbol     % make vectors bold

\newcommand{\A}{\mathrm{A}}
\newcommand{\B}{\mathrm{B}}

\begin{document}

%%%%%%%%%%%%%%%%%%%%%%%%%%%%%%%%%%%%%%%%%%%%%%%%%%%%%%%%%%%%%
%%%%%%%%%%%%%%%%%%%%%%%%%%%%%%%%%%%%%%%%%%%%%%%%%%%%%%%%%%%%%
%%%%%%%%%%%%%%%%%%%%%%%%%%%%%%%%%%%%%%%%%%%%%%%%%%%%%%%%%%%%%

\title{Comment on ``Skyrmion Hall Effect in Altermagnets"}

\author{Nicolai Timon Bechler}
\email{nicolai.bechler@kit.edu}
\author{Jan Masell}%
\email{jan.masell@kit.edu}
\affiliation{
Institute of Theoretical Solid State Physics, Karlsruhe Institute of Technology, Karlsruhe, 76131, Germany
}
%\maketitle

% Hack for title in PRL style
\noindent\textbf{Comment on ``Skyrmion Hall Effect in Altermagnets"}\\

%%%%%%%%%%%%%%%%%%%%%%%%%%%%%%%%%%%%%%%%%%%%%%%%%%%%%%%%%%%%%
%%%%%%%%%%%%%%%%%%%%%%%%%%%%%%%%%%%%%%%%%%%%%%%%%%%%%%%%%%%%%
%%%%%%%%%%%%%%%%%%%%%%%%%%%%%%%%%%%%%%%%%%%%%%%%%%%%%%%%%%%%%

%%%%%%%%%%%%%%%%%%%%%%%%%%%%%%%%%%%%%%%%%%%%%%%%%%%%%%%%%%%%%
%%%%%%%%%%%%%%%%%%%%%%%%%%%%%%%%%%%%%%%%%%%%%%%%%%%%%%%%%%%%%
% Introduction
% What have they done?
% SkHE is observed in simulations and explained in formulas using quadrupole moment.
% BUT: Dimensionless quadrupole tensor elements A_ij = 0.

In their Letter, Jin \emph{et al.}~\cite{Jin2024} predict a transverse velocity $v_y$ for skyrmions driven by spin-transfer torque (STT), known as skyrmion Hall effect (SkHE), in $d$-wave altermagnets.
Usually, the SkHE emerges from the skyrmion's topological charge which here cancels between sublattices.
Instead, Jin \emph{et al.} attribute the SkHE to the dimensionless quadrupole tensor $\mathcal{A}_{i\!j}\!=\!\int \partial_j \! \left(\vec{\mathcal{A}} \!\cdot\! \partial_i \vec{l}\right) \mathrm{d}^2r$, $\vec{\mathcal{A}} \!=\!\frac{A_2}{A_0} (\partial_x^2\vec{l} - \partial_y^2\vec{l}) \!\times\! \vec{l}$.
The result $v_y = \mathcal{A}_{xy}u_x / (\alpha \mathcal{D}_{yy})$, their Eq.~(12), is fitted to micromagnetic simulations yielding $\mathcal{A}_{xy}=-2.808\times10^{-3}$.
However, via the divergence theorem, $\mathcal{A}_{i\!j}$ simplifies to a boundary integral
\begin{equation}
    \mathcal{A}_{i\!j}
    % repeat original formula
    %= \int_\Omega \partial_j \left(\vec{\mathcal{A}} \cdot \partial_i \vec{l}\right) \mathrm{d}^2r
    % rewrite as divergence of a vector whose non-j components are zero
    %= \int_\Omega \sum_{k}\partial_k \left(\delta_{jk} \left(\vec{\mathcal{A}} \cdot \partial_i \vec{l}\right)\right) \mathrm{d}^2r
    % use divergence/Gauss's theorem: integrate over surface d\vec{s}, scalar product with the vector.
    %= \oint_{\partial\Omega} \sum_{k} \left(\delta_{jk} \left(\vec{\mathcal{A}} \cdot \partial_i \vec{l}\right)\right) \mathrm{d}s_k
    % use that k=j but remember that ds is the surface vector, pointing outward.
    % Example: Let j=x, then while integrating along x, ds is along y, perpendicular to the auxiliary vector. 
    %= \oint_{\partial\Omega} \left(\vec{\mathcal{A}} \cdot \partial_i \vec{l}\right) \mathrm{d}s_j
    % or write like this
    = \oint_{\partial\Omega} \left(\vec{\mathcal{A}} \cdot \partial_i \vec{l}\right)\hat{n}_j \mathrm{d}s
    =0    
\end{equation}
that vanishes for a sufficiently smooth and localized texture.
In this Comment, we substantiate our claim from Ref.~\cite{Bechler2026} that the numerically observed SkHE by Jin \emph{et al.} for a micromagnetic model with standard Zhang-Li STT in the individual sublattices is merely a discretization artefact.

%%%%%%%%%%%%%%%%%%%%%%%%%%%%%%%%%%%%%%%%%%%%%%%%%%%%%%%%%%%%%
%%%%%%%%%%%%%%%%%%%%%%%%%%%%%%%%%%%%%%%%%%%%%%%%%%%%%%%%%%%%%
% Equivalent m-picture description:
% (a) Introduce the coupled m description.
% (b) Mention symmetries.
% (c) Solve to show zero SkHE.

Instead of the N\'eel vector $\vec{l}=(\vec{m}^\A-\vec{m}^\B)/2$, we employ the magnetization fields $\vec{m}^{i}$ on their respective sublattices $i \in \left\{\A, \B\right\}$, like the numerical implementation.
The Thiele approach for the skyrmion velocities $\dot{\vec{R}}^i$ in this convention yields
\begin{align}
    &(\mathcal{G}^{i} + \alpha \mathcal{D}^{i}) \cdot \dot{\vec{R}}^i = \vec{F}^i 
    \label{eq:1}
    \\
    &\text{with } \vec{F}^i = \vec{F}^i_\text{coupl.} + \left( {\mathcal{G}}^{i} + \beta \mathcal{D}^{i} \right) \cdot \vec{u}
    \label{eq:2}
    \,\,.
\end{align}
The attractive force between $\vec{R}^\A$ and $\vec{R}^\B$ is $\vec{F}^\A_\text{coupl.} = -\vec{F}^\B_\text{coupl.}$.
$\mathcal{G}^{i}$ and $\mathcal{D}^{i}$ are the gyromagnetic and dissipation tensors, respectively~\cite{Iwasaki2013}, which in a $d$-wave altermagnet obey $\mathcal{G}^\A+\mathcal{G}^\B=0$ and $\mathcal{D}^\A+\mathcal{D}^\B=2\mathcal{D}_{0}\mathds{1}$ with the scalar $\mathcal{D}_{0} = (\mathcal{D}^\A_{xx} + \mathcal{D}^\A_{yy})/2 = (\mathcal{D}^\B_{xx} + \mathcal{D}^\B_{yy})/2$.
In the steady state $\dot{\vec{R}}^\A=\dot{\vec{R}}^\B\equiv\dot{\vec{R}}$, 
%the coupled Thiele equations 
Eqs.~\eqref{eq:1} and \eqref{eq:2} yield
\begin{equation}
    \dot{\vec{R}} 
    = \frac{1}{2\alpha\mathcal{D}_{0}} \left( \bm{F}^\mathrm{A} + \bm{F}^\mathrm{B} \right)
    = \frac{\beta}{\alpha} \vec{u}
    \,\,.
\end{equation}
Hence, the SkHE vanishes, consistent with $\mathcal{A}_{i\!j}=0$.

%%%%%%%%%%%%%%%%%%%%%%%%%%%%%%%%%%%%%%%%%%%%%%%%%%%%%%%%%%%%%
%%%%%%%%%%%%%%%%%%%%%%%%%%%%%%%%%%%%%%%%%%%%%%%%%%%%%%%%%%%%%
% How to explain numerics:
% (a) Remind of how STT is implemented.
% (b) Derive HOSTT.
% (c) Derive HOSTT correction to Thiele.
% (d) Solve.

\begin{figure}[t]
    \centering
    \includegraphics[width=0.95\linewidth]{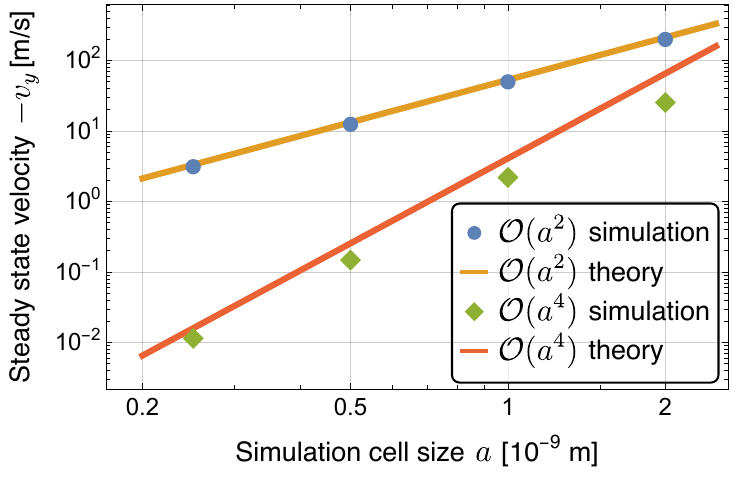}
    \caption{
    Velocity $-v_y$ as a function of simulation cell size $a=a_x=a_y$.
    $\mathcal{O}(a^n)$ simulations use $\mathcal{O}(a^n)$ finite difference schemes.
    Curves are the corresponding Thiele solutions with higher order STT.
    The simulation parameters are 
    $M_s = 290 \, \mathrm{kA}/\mathrm{m}$, 
    $A_1 = 40 \, \mathrm{pJ}/\mathrm{m}$, 
    $|A_2| = 16 \, \mathrm{pJ}/\mathrm{m}$, 
    $A_0 = 2 \, \mathrm{MJ}/\mathrm{m^3}$, 
    $D = -4 \, \mathrm{mJ}/\mathrm{m^2}$, 
    $K = 800 \, \mathrm{kJ}/\mathrm{m^3}$, 
    $\alpha = 0.01$,
    $\beta = 0.02$ and 
    $u = 200 \, \mathrm{m}/\mathrm{s}$.
    }
    \label{fig:1}
\end{figure}

Now, we explain its origin in numerical simulations.
The STT is proportional to $(\vec{u}\!\cdot\!\nabla)\vec{m}^i$.
In MuMax3~\cite{Vansteenkiste2014}, a derivative $\partial_{\nu} \vec{m}^i(\vec{r})$ is implemented as a finite difference
\begin{equation}
\begin{split}
    \frac{1}{2 a_\nu}&\big(\vec{m}^i\left( \vec{r} + a_\nu \hat{e}_\nu \right) - \vec{m}^i\left( \vec{r} - a_\nu \hat{e}_\nu \right) \big) \\
    \approx 
    \,&\partial_{\nu}\vec{m}^i(\vec{r}) 
    + 
    \bigg[ \frac{a_\nu^2}{6} \partial^3_{\nu}\vec{m}^i(\vec{r}) + \mathcal{O}\left( a^4_\nu \right) \bigg]
\end{split}
\label{eq:4}
\end{equation}
with mesh size $a_\nu$ and unit vectors $\hat{e}_\nu$, $\nu=x,y,z$.
The part in square brackets is usually neglected.
However, starting from the numerical implementation, the $\mathcal{O}(a_\nu^2)$ terms indeed appear in the LLG equation of motion, adding higher order STT $u_\nu\partial_{\nu}\vec{m}^i \rightarrow u_\nu\partial_{\nu}\vec{m}^i + (a_\nu^2/6) u_\nu\partial^3_{\nu}\vec{m}^i + \mathcal{O}( a^4_\nu )$.
These additional torques straightforwardly translate into the Thiele  equation, modifying in Eq.~\eqref{eq:2} the force due to the current as
$\mathcal{G}^i \rightarrow \sum_{\substack{n=1 \\ n \text{ odd}}}^\infty \mathcal{G}^{(n), i}$ 
and 
$\mathcal{D}^i\rightarrow \sum_{\substack{n=1 \\ n \text{ odd}}}^\infty \mathcal{D}^{(n),i}$:
\begin{align}
    \mathcal{G}^{(n), i}_{\mu \nu} 
    %&(-1)^{n} \frac{a^{n-1}_\nu}{n!} \int \bm{m}^i \cdot \left( \partial_\mu \bm{m}^i \right) \times \left( \partial^n_\nu \bm{m}^i \right) \mathrm{d}^2 r
    &= - \frac{a^{n-1}_\nu}{n!} \int \bm{m}^i \cdot \left( \partial_\mu \bm{m}^i  \times  \partial^n_\nu \bm{m}^i \right) \mathrm{d}^2 r
    \label{eq:5}
    \\
    \mathcal{D}^{(n),i}_{\mu \nu} 
    %&(-1)^{n-1} \frac{a^{n-1}_\nu}{n!} \int \left( \partial_\mu \bm{m}^i \right) \cdot \left( \partial^n_\nu \bm{m}^i \right) \mathrm{d}^2 r
    &=\phantom{-} \frac{a^{n-1}_\nu}{n!} \int  \partial_\mu \bm{m}^i \cdot  \partial^n_\nu \bm{m}^i \mathrm{d}^2 r
    \label{eq:6}
    \,\,,
\end{align}
where $n$ is the order of the derivative $\partial_\nu^n$ in Eq.~\eqref{eq:4}.
We then obtain
\begin{equation}
    v_y 
    \approx
    \left( 
    \mathcal{G}^{(3),\A}_{yx} + 
    \mathcal{G}^{(3),\B}_{yx} + 
    \left( \mathcal{D}^{(3),\A}_{yx} + \mathcal{D}^{(3),\B}_{yx} \right) \beta
    \right)
    \frac{ u_x }{ 2 \mathcal{D}_0 \alpha }
    ,
    \label{eq:8}
\end{equation}
where the tensor elements all scale with the numerical lattice constant as $a_\nu^2$.
We find agreement in the small mesh regime between this analytical result and numerical scaling in Fig.~\ref{fig:1} using MuMax3~\cite{Vansteenkiste2014}, confirmed with $\text{mumax}^+$~\cite{Moreels2026}.
We also include results for an $\mathcal{O}(a^4)$ numerical implementation~\cite{Bechler2026}, where the five-point-stencil version of Eq.~\eqref{eq:4} has a leading correction $-(a^4_\nu/30)\partial_\nu^5\vec{m}^i$, leading again to Eq.~\eqref{eq:8} but with the replacement $\mathcal{T}^{(3),\A/\B}_{yx} \to -4\mathcal{T}^{(5),\A/\B}_{yx}$, $\mathcal{T}=\mathcal{G},\mathcal{D}$.

In conclusion, the approach in Ref.~\cite{Jin2024} via $\mathcal{A}_{xy}$ cannot generate a SkHE for sufficiently localized textures.
In micromagnetic simulations, it stems from higher order STT introduced by the numerical discretization.
Our theory with this higher order STT explains the numerical results and predicts a vanishing SkHE for $a_\nu\to0$.

%%%%%%%%%%%%%%%%%%%%%%%%%%%%%%%%%%%%%%%%%%%%%%%%%%%%%%%%%%%%%
%%%%%%%%%%%%%%%%%%%%%%%%%%%%%%%%%%%%%%%%%%%%%%%%%%%%%%%%%%%%%
% Acknowledgement

\textit{Acknowledgements} --- We acknowledge funding from the Deutsche Forschungsgemeinschaft (DFG, German Research Foundation) under Project No. 547968854.

%%%%%%%%%%%%%%%%%%%%%%%%%%%%%%%%%%%%%%%%%%%%%%%%%%%%%%%%%%%%%
%%%%%%%%%%%%%%%%%%%%%%%%%%%%%%%%%%%%%%%%%%%%%%%%%%%%%%%%%%%%%
% Author list

\vspace{1em}

\noindent
N. T. Bechler\,\href{https://orcid.org/0009-0006-6824-3537}{\includegraphics[height=0.75em]{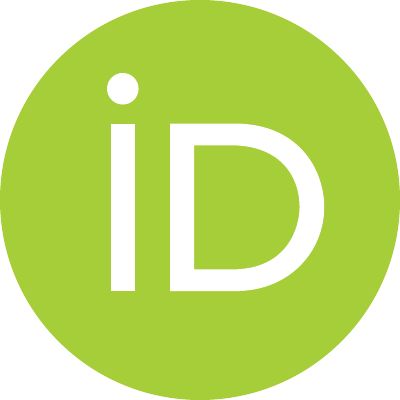}} 
and 
J. Masell\,\href{https://orcid.org/0000-0002-9951-4452}{\includegraphics[height=0.75em]{orcid.pdf}}$^\ast$

{\small Institute of Theoretical Solid State Physics}
\vspace{-0.25em}

{\small Karlsruhe Institute of Technology}
\vspace{-0.25em}

{\small 76131 Karlsruhe, Germany}

\vspace{0.25em}
$^\ast$ Contact author: \href{mailto:jan.masell@kit.edu}{jan.masell@kit.edu}

\newpage

%%%%%%%%%%%%%%%%%%%%%%%%%%%%%%%%%%%%%%%%%%%%%%%%%%%%%%%%%%%%%
%%%%%%%%%%%%%%%%%%%%%%%%%%%%%%%%%%%%%%%%%%%%%%%%%%%%%%%%%%%%%
% bibliography

\bibliography{bibliography}

\end{document}